\pdfoutput=1
\documentclass[prd,preprint,superscriptaddress,preprintnumbers,nofootinbib]{revtex4}
\usepackage{graphicx}
\usepackage{epsfig}
\usepackage{bm}
\usepackage{subfigure}
\usepackage{latexsym,amssymb,amsmath,amsfonts,amssymb,wasysym,float}
\usepackage{mathrsfs}
\usepackage{color}
\usepackage{lettrine}
\usepackage{lipsum}
\usepackage{enumitem}

\def\xe{{\epsilon}}

\def\be{\begin{equation}}
\def\ee{\end{equation}}
\def\tfrac#1#2{{\textstyle\frac{#1}{#2}}}
\def\thalf{\tfrac{1}{2}}
\def\beas{\begin{eqnarray*}}
\def\eeas{\end{eqnarray*}}
\def\bea{\begin{eqnarray}}
\def\eea{\end{eqnarray}}

\def\mpl{M_{\rm Pl}}

\def\sech{\mathrm{sech}}
\def\asech{\mathrm{arcsech}}

\newcommand{\lb}{\left(}
\newcommand{\rb}{\right)}

\usepackage[usenames,dvipsnames]{xcolor}
\definecolor{orange}{cmyk}{0,0.5,1,0}
\definecolor{rossoCP3}{cmyk}{0,.88,.77,.40}
\definecolor{graa}{rgb}{0.8,0.8,0.8}
\definecolor{blaa}{rgb}{0.2,0.2,0.6}

\begin{document}
\preprint{MPP-2021-30}
\preprint{LMU-ASC 06/21}

\title{\color{rossoCP3}   $\boldsymbol{S}$-dual Inflation and the String
  Swampland}

\author{Luis A. Anchordoqui}

\affiliation{Department of Physics and Astronomy,  Lehman College, City University of
  New York, NY 10468, USA
}

\affiliation{Department of Physics,
 Graduate Center, City University
  of New York,  NY 10016, USA
}

\affiliation{Department of Astrophysics,
 American Museum of Natural History, NY
 10024, USA
}

\author{Ignatios Antoniadis}
\affiliation{Laboratoire de Physique Th\'eorique et Hautes \'Energies - LPTHE
Sorbonne Universit\'e, CNRS, 4 Place Jussieu, 75005 Paris, France
}

\affiliation{Institute for Theoretical Physics, KU Leuven, Celestijnenlaan 200D, B-3001 Leuven, Belgium
}

\author{Dieter\nolinebreak~L\"ust}

\affiliation{Max--Planck--Institut f\"ur Physik,
 Werner--Heisenberg--Institut,
80805 M\"unchen, Germany
}

\affiliation{Arnold Sommerfeld Center for Theoretical Physics
Ludwig-Maximilians-Universit\"at M\"unchen,
80333 M\"unchen, Germany
}

\author{Jorge F. Soriano}

\affiliation{Department of Physics and Astronomy,  Lehman College, City University of
  New York, NY 10468, USA
}
\affiliation{Department of Physics,
 Graduate Center, City University
  of New York,  NY 10016, USA
}

\begin{abstract}
  \vskip 2mm
\noindent  The Swampland de Sitter conjecture in combination with
  upper limits on the tensor-to-scalar ratio $r$ derived from
  observations of the cosmic microwave background endangers the
  paradigm of slow-roll single field inflation. This conjecture
  constrains the first and the second derivatives of the inflationary
  potential in terms of two ${\cal O} (1)$ constants $c$ and $c'$. In
  view of these restrictions we reexamine single-field inflationary potentials with $S$-duality
  symmetry, which ameliorate the unlikeliness problem of  the initial condition. We compute $r$ at next-to-leading order in
  slow-roll parameters for the most general form of $S$-dual
  potentials and confront model predictions to constraints imposed by the de Sitter
  conjecture. We find that $c \sim {\cal O} (10^{-1})$ and $c' \sim
  {\cal O} (10^{-2})$ can accommodate the 95\% CL upper limit on $r$. By imposing
  at least 50 $e$-folds of inflation with
    the effective field theory description only valid over a field
    displacement ${\cal O} (1)$ when measured as a distance in the target
    space geometry, we further restrict $c \sim {\cal O} (10^{-2})$, while the constraint on $c'$ remains unchanged. We comment on how
    to accommodate the required small values of $c$ and $c'$.
\end{abstract}

\maketitle

\section{Introduction}

Inflation is the leading paradigm for explaining the behavior of the
quasi-de Sitter expansion in the very early
universe. Single-field inflationary models
provide promising explanations to the cosmological horizon problem, the lack of
topological defects, and the observed large-scale isotropy~\cite{Guth:1980zm,Linde:1981mu,Albrecht:1982wi}. In
addition, inflation provides a mechanism for generating small
fluctuations in energy density, which could have seeded galactic structure
formation~\cite{Mukhanov:1981xt,Hawking:1982cz,Guth:1982ec,Starobinsky:1982ee,Bardeen:1983qw}, and are observed in the temperature
anisotropies of the cosmic microwave
background (CMB)~\cite{Spergel:2003cb,Komatsu:2010fb}.

One of the main goals of modern CMB missions is to measure the
tensor-to-scalar ratio $r$ accurately to constrain inflationary
models. The combination of BICEP2/Keck Array data with observations by
Planck (TT,TE,EE +lowE+lensing) and BAO  significantly shrink the space of allowed inflationary cosmologies:
$r< 0.068$ at 95\% CL~\cite{Ade:2018gkx,Akrami:2018odb}. Moreover, CMB data favor standard slow-roll
single field inflationary models with plateau-like potentials
$V$, for which $V_{\phi \phi} <0$, over power-law
potentials; here $\phi$ is the dilaton/inflaton and $V_\phi \equiv
dV/d\phi$~\cite{Ade:2015lrj}. In this paper we investigate slow-roll inflationary models within the context of the Swampland
program~\cite{Vafa:2005ui} and confront model predictions with
experiment. We particularize the investigation to inflationary
potentials satisfying $V_{\phi \phi} <0$ while being invariant under the
$S$-duality constraint, $ \phi \rightarrow -
\phi$~\cite{Montonen:1977sn}, which is reminiscent of String
Theory~\cite{Font:1990gx,Sen:1994fa}.

The Swampland program has been established to lay out a connection between quantum gravity and very-large-scale/ultra-low-energy  astronomical observations. The String Swampland comprises the set of (apparently) consistent
effective field theories (EFT) that cannot be completed into quantum
gravity in the ultraviolet~\cite{Brennan:2017rbf,Palti:2019pca}. This
rather abstract concept implies that if
gravity were to be added into an EFT which is self-consistent up to a scale
$E_{\rm self}$, then the combined theory would exhibit  a new limitting energy
scale $E_{\rm swamp}$, above which the theory must be modified if it is
to become compatible with quantum gravity in the ultraviolet. When the
energy relation $E_{\rm swamp} < E_{\rm self} < E_{\rm Pl}$ holds and  $E_{\rm swamp}$ is below any characteristic energy scale involved in
the theory, we say that the entire EFT belongs to the String
Swampland; $E_{\rm Pl}$ denotes the Planck energy scale. Guidance for a model building approach is provided by an
ensemble of Swampland conjectures~\cite{ArkaniHamed:2006dz,Ooguri:2006in,Klaewer:2016kiy,Ooguri:2018wrx,Grimm:2018ohb,Heidenreich:2018kpg,Ooguri:2016pdq,Palti:2017elp,Obied:2018sgi,Andriot:2018wzk,Kehagias:2018uem,Cecotti:2018ufg,Klaewer:2018yxi,Heckman:2019bzm,Lust:2019zwm,Bedroya:2019snp,Kehagias:2019akr,Blumenhagen:2019vgj,Andriot:2020lea,Gendler:2020dfp,Bonnefoy:2020uef,Perlmutter:2020buo,Luben:2020wix,Calderon-Infante:2020dhm,Kolb:2021xfn}. There are two consequential conjectures which gathered immediate interest in the context of inflationary cosmology:
\begin{itemize}[noitemsep,topsep=0pt]
\item Distance Swampland conjecture: This conjecture limits the field
  space of validity of any EFT by limiting the field excursion $\Delta
  \phi$ to be small when expressed in Planck units, namely $\Delta \phi/M_{\rm Pl}
  \equiv \delta \alt \alpha \sim {\cal O} (1)$, where $M_{\rm Pl} =
(8 \pi G)^{-1/2}$ is the reduced Planck mass~\cite{Ooguri:2006in,Klaewer:2016kiy, Ooguri:2018wrx,Grimm:2018ohb,Heidenreich:2018kpg}.
\item de Sitter conjecture: The gradient of the scalar potential $V$
  must satisfy the lower bound,
\begin{equation}
  M_{\rm Pl} \frac{|V_\phi|}{V} \equiv {\cal C} \geq c \,,
\label{dS1}
\end{equation}
  or else its Hessian must satisfy
\begin{equation}
  M_{\rm Pl}^2 \frac{V_{\phi \phi}}{V} \equiv {\cal C'} \leq - c'
\label{dS2}
\end{equation}
where $c$ and
$c'$ are positive order-one numbers in Planck units~\cite{Obied:2018sgi,Ooguri:2018wrx}.
\end{itemize}
It has been noted that while  the distance conjecture by itself does not pose
significant challenge for single-field inflationary models (and
corresponds observationally to a suppressed $r$ through the well-known
Lyth bound~\cite{Lyth:1996im}), the de Sitter conjecture is in direct tension with
slow-roll inflationary potentials favored by CMB
data~\cite{Agrawal:2018own,Achucarro:2018vey,Garg:2018reu,Ben-Dayan:2018mhe,Kinney:2018nny,Fukuda:2018haz,Garg:2018zdg,Agrawal:2018rcg,Chiang:2018lqx}.  The
objective of our investigation is to analyze the status of  single-field  inflationary  potentials  with $S$-duality symmetry in the context of the Swampland conjectures.

The remainder of the paper is structured as follows. In
Sec.~\ref{sec:2}  we first provide an overview of  the equations of motion
in single-field slow-roll inflation and introduce  the definition of
the slow-roll parameters. After that, to make the connection with experiment we compute the  scalar spectral index $n_s$
and the tensor-to-scalar ratio $r$ at next-to-leading order (NLO) in
slow-roll parameters.  In
Sec.~\ref{sec:3} we examine the subtleties of model building while
imposing contraints which depend on multiple slow-roll
parameters focussing attention on $S$-dual symmetric inflationary
potentials. We summarize the generalities of these potentials and confront model
predictions to the  CMB
observables $n_s$ and $r$. In Sec.~\ref{sec:4-new} we investigate the ambiguity on the definition of the slow-roll parameters and explore whether this uncertainty can help ameliorate the tension between single-field inflationary models and the de Sitter Swampland conjecture. The paper wraps up with some conclusions presented in  Sec.~\ref{sec:4}.

\section{Constraints on $\boldsymbol{r}$ at NLO in slow-roll parameters}
\label{sec:2}

The essential property of nearly all crowned inflationary models is a
period of slow-roll evolution of $\phi$ during which its kinetic
energy remains always much smaller
than its potential energy. The equation of motion for the canonical
homogeneous inflaton field is
\be
\ddot{\phi} + 3 H\dot{\phi} + V_\phi = 0 \,,\label{eq:eom}
\ee
where $H = \dot a/a$ is the Hubble parameter and dot denotes derivative with respect
to the cosmic time. The slow-roll conditions
\be
\thalf \dot{\phi}^2 \ll  |V|
\ee
and
\be
\left| \frac{\ddot{\phi}} {3 H \dot{\phi}}\right|\ll 1  \,,
\ee
imply
\be
\epsilon \equiv  \frac{\mpl^2}{2} \left(\frac{ V_\phi}{V}\right)^2\ll
1
\ee
and
\be
\eta       \equiv  \mpl^2 \left[ \frac{ V_{\phi \phi}} {V} -
  \frac{1}{2} \left(\frac{V_\phi}{V} \right)^2\right] \ll 1 \,,
\ee
respectively~\cite{Kolb:1994ur,Dodelson:1997hr}.\footnote{The definition of the slow-roll parameters vary; we follow the conventions of~\cite{Lidsey:1995np}.} The Friedmann relation incorporating slow-roll is given by
\be
H(\phi)\simeq \sqrt{\frac{V}{3\mpl^2}} \, .
\ee
At the end of slow roll, $\phi$ falls into the core of the potential and oscillates
rapidly around the minimum, ultimately leading to the reheating
period. The amount of inflationary expansion within a given timescale is generally parametrized in terms of the number of $e$-foldings that
  occur as the scalar field rolls from a particular value $\phi$ to
  its value $\phi_e$ when inflation ends:
  \begin{equation}
    N (\phi \to \phi_e) = - \frac{1}{M_{\rm Pl}^2} \int_\phi^{\phi_e} \frac{V}{V_\phi}
    d\phi \, ,\label{eq:nefolds}
  \end{equation}
  with $\xe(\phi_e)=1$~\cite{Remmen:2014mia}. The de Sitter conjecture bounds the integrand above.
  Around a minimum of the potential without changes in the curvature
    \begin{equation}
      N(\phi\to\phi_e)=\frac{1}{\mpl^2}\int_{\min(\phi,\phi_e)}^{\max(\phi,\phi_e)}\left|\frac{V}{V_\phi}\right|d\phi\leq  \frac{|\phi_e-\phi|}{\mpl^2}\max_{[\phi,\phi_e]}\left|\frac{V}{V_\phi}\right|=\frac{|\phi_e-\phi|}{\mpl^2}\left|\frac{V}{V_\phi}\right|_{\phi},
  \end{equation} as $V/V_\phi\sim1/\sqrt\epsilon$ grows as the field moves away from the minimum. Using the de Sitter conjecture bound this can be written more compactly as~\cite{Fukuda:2018haz} \begin{equation}\frac{\Delta\phi}{\mpl} >c\,N.\label{eq:delta_bound}\end{equation}

 To make contact with experiment we calculate $r$ at NLO in slow-roll parameters. We begin by parametrizing the scalar
            \begin{equation} {\cal
  P}_s  =  A_s \lb \frac k {k_*} \rb^{n_s -1 + \frac 1 2
  \alpha_s  \ln \lb \frac k {k_*} \rb  + \cdots}
\end{equation}
and tensor
\begin{equation}
{\cal P}_t  =  A_t \lb \frac k {k_*} \rb^{n_t + \frac 1 2 \alpha_t
  \ln \lb \frac k {k_*} \rb + \cdots}\,
\end{equation}
 power spectra, where the spectral indices and their running (included here for completeness only) are given by
\begin{eqnarray}
n_s& \simeq & 1-4\epsilon+2\eta+ \left(\frac {10} 3 +4\mathcal B \right) \xe
\eta - (6+4\mathcal B ) \xe^2 + \frac 2 3 \eta^2 - \frac{2}{3} (3 \mathcal B -1) \left(2 \xe^2-6 \xe \eta +\xi ^2\right)\,, \nonumber \\
n_t &\simeq & -2\epsilon + \left(\frac 8 3 +4\mathcal B  \right) \xe \eta -\frac
              2 3 (7+6\mathcal B ) \xe^2 \,,  \nonumber \\
  \alpha_s & \equiv & {\frac{d n_s}{d \ln k}}  \simeq  -8 \epsilon ^2+
16 \epsilon \eta -2\xi^2\,,  \nonumber \\
\alpha_t & \equiv & \frac{d n_t}{d \ln k}  \simeq  -4\epsilon(\epsilon-\eta)\,,
\label{eqs:indices}
\end{eqnarray}
and where $\mathcal B  = \gamma_{\rm E} +
                    \ln 2 - 2 \approx
                    -0.7296$ and
\be
\xi^2  \equiv  \frac {M_{\rm Pl}^4 V_\phi V_{\phi \phi \phi}} {V^2} \,,
\label{xi2}
\ee
is the third slow-roll parameter~\cite{Leach:2002ar}. The NLO amplitudes are related to $\xe,\eta$ and $V$ by
\bea
  A_s &\simeq & \frac{V}{24 \pi^2 M_{\rm{Pl}}^4\epsilon}\left[1 - (4\mathcal B  + 1)
    \epsilon + \lb 2\mathcal B  -\frac 2 3\rb\eta\right]\,, \label{eq:amplitude} \\
A_t & \simeq &   \frac{V}{6 \pi^2 M_{\rm{Pl}}^4}\left[1 - \left(2\mathcal B  + \frac 5
    3 \right) \epsilon \right]  \, .
\eea
All in all, the ratio of the NLO amplitudes of the spectra is given by
\be
   r\equiv \frac{A_t}{A_s} \simeq 16 \epsilon + 32\lb \mathcal B  - \frac 1 3 \rb \xe (\xe - \eta) \, .
\label{eq:r}
 \ee

\begin{figure}
    \centering
    \subfigure{\includegraphics[height=.465\linewidth]{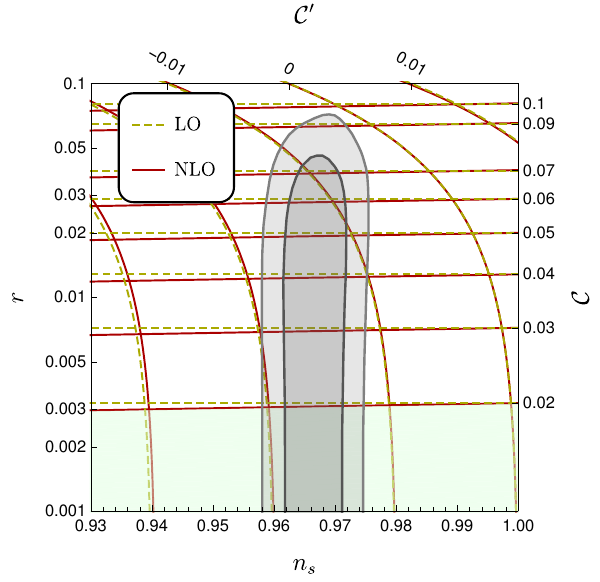}}
    \subfigure{\includegraphics[height=.4\linewidth]{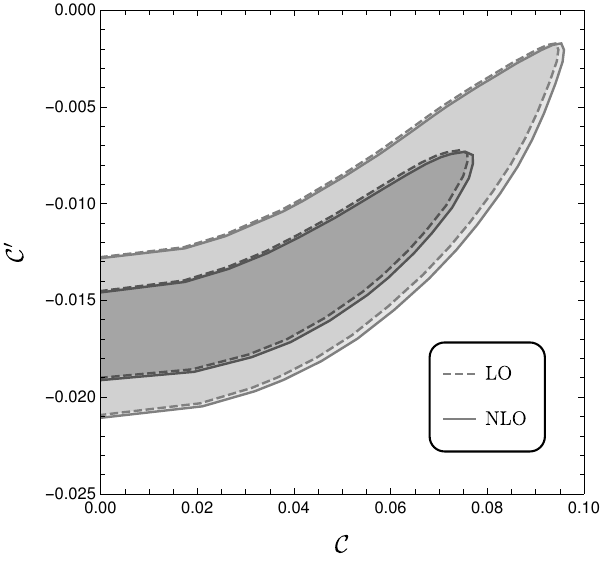}}
    \caption{Relation between $(r,n_s)$ and $(\mathcal C,\mathcal C')$ at LO and NLO, together with the bound (\ref{eq:delta_bound}) (shaded region) for $N=50$ and $\Delta\phi=\mpl$ (left); and experimental constrains on ($\mathcal C,\mathcal C')$ at LO and NLO from TT, TE, EE + lowE + lensing + BK15 + BAO data~\cite{Akrami:2018odb} (right).}
    \label{fig:nlo}
\end{figure}

Substituting (\ref{dS1}) into (\ref{eq:r}) we reproduce the
well-known constraint at LO in slow-roll parameters,
\begin{equation}
  r \simeq 16 \epsilon \equiv  8 {\cal C}^2 \Rightarrow {\cal C} \simeq \sqrt{r/8} \alt
  0.09 \, ,
\label{eq:upc}
\end{equation}
where we have taken $r$ to saturate the 95\% CL upper limit.  A  comparison of this upper limit with the lower limit in (\ref{dS1}) has called
  into question whether slow-roll single field inflationary models
 could live on the
 Swampland~\cite{Agrawal:2018own,Achucarro:2018vey,Garg:2018reu,Ben-Dayan:2018mhe,Kinney:2018nny,Fukuda:2018haz,Garg:2018zdg,Agrawal:2018rcg,Chiang:2018lqx}.  Using the upper value of the measured $1\sigma$ range, $n_s = 0.9658
 \pm 0.0040$~\cite{Akrami:2018odb}, a
 combined limit on  ${\cal C}$ and ${\cal C}'$ can be derived substituting
(\ref{dS1}) and (\ref{dS2}) into the expression of the scalar
spectral index (\ref{eqs:indices}). At LO,
\begin{equation}
n_s \simeq 1 - 2 {\cal C}^2 + 2 \eta \, .
\end{equation}
The allowed region of the $n_s-r$ plane at LO has been reported
in~\cite{Kinney:2018nny}. We can visualize the modification of the NLO
bounds on ${\cal C}$ and ${\cal C}'$ posed by the data in Fig.~\ref{fig:nlo} in a model independent way, to the degree that the $\xi^2$ term in the expansion for $n_s$ in (\ref{eqs:indices}) is negligible. This alone suggest a certain degree of incompatibility between observations and the de Sitter conjecture. The inclusion of a non-zero $\xi$ would slightly reduce the tension with $c'$, displacing down the contours of Fig.~\ref{fig:nlo} (right) but leaving them almost unchanged along the $\mathcal C$ direction.

In the next section we will particularize our study to inflationary  potentials with $S$-duality symmetry. In particular, we will explore the relevance of the distance Swampland conjecture, which cannot be explored in a model independent way at any order.

\section{$\boldsymbol{S}$-duality strikes again}
\label{sec:3}

Dualities within gauge theories are striking as they relate a strongly
coupled field theory to a weakly coupled one, and thereby they are handy for evaluating a theory at strong coupling, where
perturbation theory breaks down, by translating it into its dual
description with a weak coupling constant; ergo, dualities point
to a single quantum system which has two classical limits. The $U(1)$
gauge theory on $\mathbb R^4$ is known to possess an electric-magnetic
duality symmetry that inverts the coupling constant and extends to an
action of $SL(2,\mathbb Z)$~\cite{Montonen:1977sn}. There are also many examples of $S$-duality in String Theory~\cite{Font:1990gx,Sen:1994fa}. In this section we examine potentials which are invariant under the
$S$ duality constraint and confront them with experiment. Herein
we do not attempt a full association with a particular string vacuum,
but simply regard the self-dual constraint as a relic of string
physics in inflationary cosmology. We adopt a  phenomenological
approach to expand the inflationary potential in terms of a generic
form satisfying the $S$-duality constraint and then the
determination of the expansion coefficients is data driven.

For a real scalar field $\phi$, the $S$-duality symmetry is $\phi \to
-\phi$  (or alternatively \mbox{$g\rightarrow 1/g$}, with $g\sim e^{\phi/M_{\rm Pl}}$). In case there is an imaginary part, i.e. an axion, then the
$S$-duality group is extended to the modular group $SL(2, \mathbb
Z)$. The $S$-duality constraint forces a
particular functional form on the inflationary potential:
$V(\phi) = f[\cosh(\kappa \phi/M_{\rm Pl})]$, where $\kappa$ is a
constant~\cite{Anchordoqui:2014uua}.

A compelling property of inflationary potentials featuring $S$-duality symmetry is that they resolve the ``unlikeliness problem'', which is
typical of  plateau-like potentials, e.g.,
\begin{equation}
  V_1(\phi) = \frac{V_0^{(1)}}{M_{\rm Pl}^4} \left(\phi^2 - \phi_0^2 \right)^2,
\label{eq:V1}
\end{equation}
where $V_0$ and $\phi_0$ are free
parameters~\cite{Ijjas:2013vea}. Note that the plateau region satisfies $\phi \ll \phi_0$ terminating at the local
minimum, and for large values of $\phi$ the
potential grows as a power-law $\sim V_0 (\phi/M_{\rm
  Pl})^4$. This means that we have two paths
to reach the minimum of the potential: by slow-roll along the plateau
or by slow-roll from the power-law side of the minimum. The problem
appears because the  path from the power-law side requires less
fine tuning of parameters, has inflation occurring over a much wider
range of $\phi$, and produces exponentially more inflation, but still
CMB data prefer the unlikely path along the plateau.

\begin{figure}
    \centering
    \subfigure{\includegraphics[clip,height=.3225\linewidth]{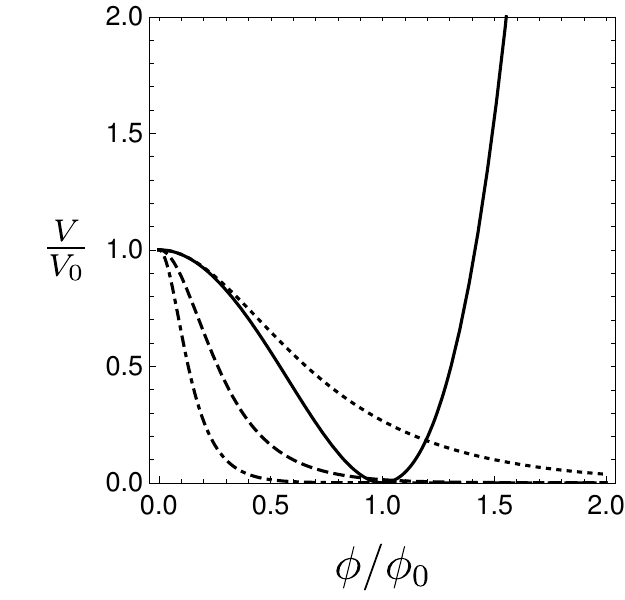}}
    \subfigure{\includegraphics[clip,height=.3225\linewidth]{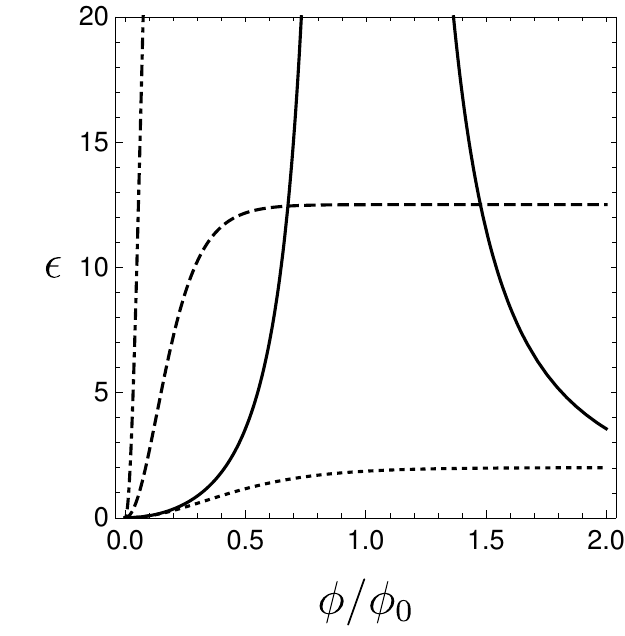}}
    \subfigure{\includegraphics[clip,height=.3225\linewidth]{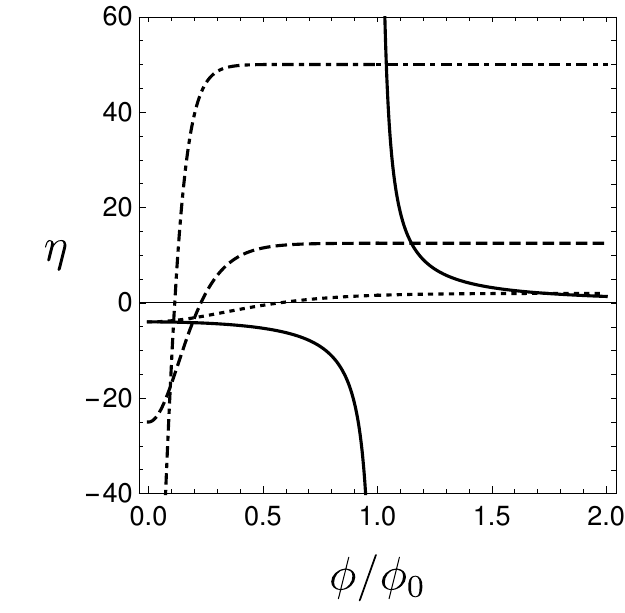}}
    \caption{Potential and slow-roll functions for the potentials defined by (\ref{eq:V1}) (solid line), and (\ref{eq:V2}) with $\kappa=2$ (dotted), $\kappa=5$ (dashed) and $\kappa=10$ (dot-dashed). \label{fig:potentials_0}}
\end{figure}

The simplest  $S$ self-dual form,
\be
V_2 (\phi) = V_0^{(2)} \ {\rm sech} \left(\frac{\kappa \phi}{M_{\rm Pl}} \right),
\label{eq:V2}
\ee
solves the unlikeliness problem because it has no power-law
wall. Moreover, it is easily seen that for (\ref{eq:V1}) and (\ref{eq:V2}) the
slow-roll parameters $\xe$ and $\eta$ are of the scale $(\phi_0/M_{\rm
  Pl})^2 \sim \kappa^{-1}$ and thus have similar inflationary growths;
see Fig.~\ref{fig:potentials_0}. However, for (\ref{eq:V1}) the slow-roll
parameters $\xe$ and $\eta$ grow fast near the end of inflation
($\phi \sim \phi_0$), but for the $S$ self-dual form $\xe$ and $\eta$
remain small because the potential has no local minimum. Thereby, $\phi$ cannot exit the
inflationary period.

To describe $S$-dual potentials for which inflation ends we adopt a polynomial  expression in the sech function. Without loss of generality, we can write it as
\begin{equation}
  V(\phi) = V_0 \sum_{n=0}^N a_n \ \sech^n \left(\frac{\kappa \phi}{M_{\rm Pl}} \right)\,,
\label{eq:nothie}
\end{equation}
under the condition that $\sum_ia_i=1$, to ensure that $V_0=V(0)$. Here, the normalization constant $V_0$ and the expansion coefficients $a_n$ are determined empirically by
matching experimental constraints. To determine the
coefficients $a_n$ we demand:
\begin{itemize}[noitemsep,topsep=0pt]
\item $N(\phi_* \to \phi_e) \simeq 60$, with $\phi_*$  the field value
  when the $k_*$ scale crosses the horizon, $k_* = a H$;
\item the NLO expression of $r$ given in (\ref{eq:r}) to satisfy the 95\% CL
  upper limit, i.e. $r < 0.069$~\cite{Akrami:2018odb};
\item the NLO expression of the scalar spectral index $n_s$ given in (\ref{eqs:indices}) to match the upper end of the measured $1\sigma$ value, $n_s \simeq 0.9698$~\cite{Akrami:2018odb}.
\end{itemize}
% The normalization constant $V_0$ is determined using the amplitude of
% the scalar primordial power spectrum given in (\ref{eq:amplitude}) in
%                combination with the best fit value to
%                Planck (TT,TE,EE+lowE+lensing) data $\ln (10^{10} A_s) =
%                3.044 \pm 0.014$~\cite{Akrami:2018odb}.

The phenomenological expression in (\ref{eq:V2}) could develop a
minimum to support dissipative oscillations at the cessation of the
slow roll and reheating, and resolves the unlikeliness problem. In
order to analyze the model, it is convenient to define
$y=\sech(\kappa\phi/\mpl)$ and $V(\phi)=V_0f(y)$,
with \begin{equation} f(y)=\sum_{n=0}^Na_n y^n. \label{eq:ref}\end{equation}

Without conflicting with $S$-duality, we restrict ourselves here to $\phi>0$ to guarantee a bijection between $y$ and $\phi$. Note that $y\in[0,1]$ as $\phi\in[0,\infty)$. It is then easy to see that
\begin{subequations}\begin{equation}V_\phi=-V_0\frac{\kappa }{\mpl}y\sqrt{1-y^2}f'(y),\end{equation}
\begin{equation}V_{\phi\phi}=V_0\left(\frac{\kappa }{\mpl}\right)^2\left[\vphantom{\sum}y^2(1-y^2)f''(y)+y(1-2y^2)f'(y)\right],\end{equation}
and
\begin{equation}V_{\phi\phi\phi}=
V_0\left(\frac{\kappa}{\mpl}\right)^3y\sqrt{1-y^2}\left[\vphantom{\sum}-y^2(1-y^2)f'''(y)-3y(1-2y^2)f''(y)-(1-6y^2)f'(y)\right] \end{equation}\label{eq:y-der}\end{subequations}
which allow to obtain analytical expressions for $\epsilon$, $\eta$ and $\xi$.

Non-trivial potentials occur for $N\geq2$. Here, we study the polynomial form in (\ref{eq:V2}) at lowest order, i.e. $f(y)=a_0+a_1 y+a_2 y^2$. From the initially four model parameters $(V_0,a_1,a_2,a_3)$ the normalization condition $V(0)=V_0$ (or $f(1)=1)$ allows to remove one of them. The potential has a minimum, at which we can impose $V=0$, removing another constant. It is easily seen that in this case $f$ can be rewritten as   \begin{equation}
    f(y)= \left(\frac{y-\beta}{1-\beta}\right)^2,\label{eq:f2}
  \end{equation}
where $\beta\in(0,1)$ is the position of the minimum. This corresponds to a potential
\begin{equation}
V(\phi)=V_0\left[\frac{\sech\left(\kappa\frac\phi\mpl\right)-\sech\left(\kappa\frac{\phi_0}{\mpl}\right)}{1-\sech\left(\kappa\frac{\phi_0}{\mpl}\right)}\right]^2,\label{eq:fullV}
\end{equation}
where
$\beta=\sech\left(\kappa\phi_0/\mpl\right)$.
A point worth noting at this juncture is that the the
  expansion of (\ref{eq:nothie}) is not hierarchical, i.e.,
the coefficients $a_n$ should not necessarily become smaller and
smaller with larger $n$. Our  choice is based on the complexity
of the model, in which larger $N$ potentials would contain more free
parameters and, under some conditions, more maxima/minima. Note that an identification of (\ref{eq:ref}) with (\ref{eq:f2})
allows one to see that $a_1/a_0=-2/\beta$ and $a_2/a_1=-1/2\beta$ and the
hierarchy, if existing, is contingent on the position of the minimum
of the field $\phi_0$ and on $\kappa$.

\begin{figure}
    \centering
    \subfigure{\includegraphics[clip,height=.3225\linewidth]{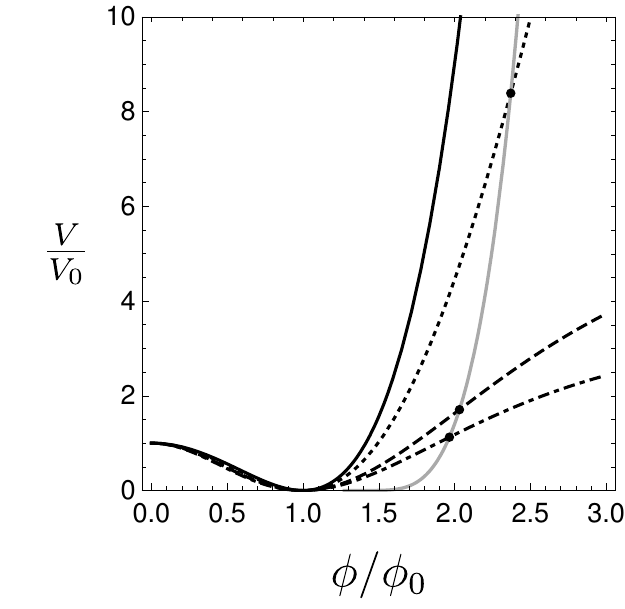}}
    \subfigure{\includegraphics[clip,height=.3225\linewidth]{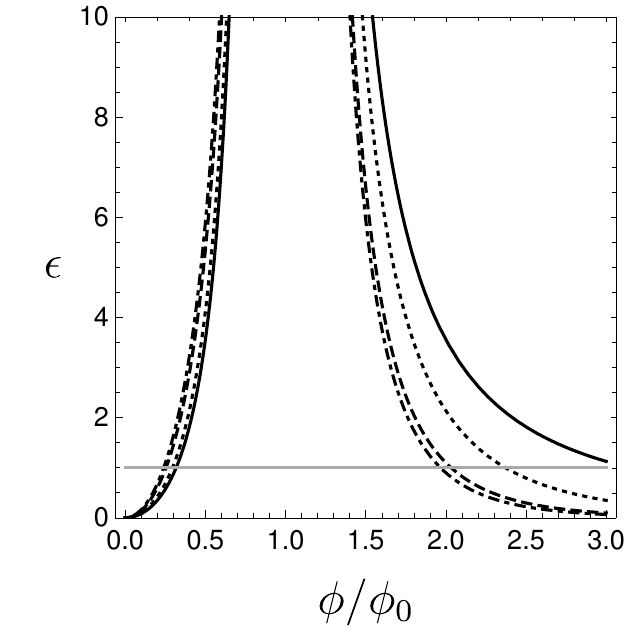}}
    \subfigure{\includegraphics[clip,height=.3225\linewidth]{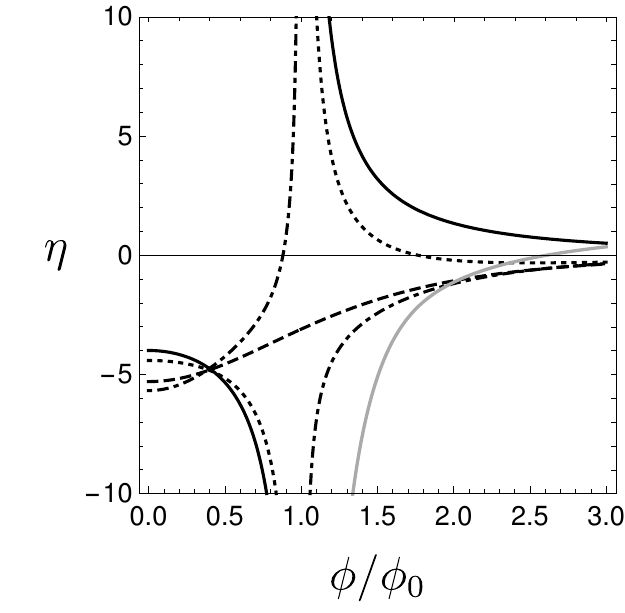}}
    \label{fig:potentials}
    \caption{Potential and slow-roll functions for the potentials defined by (\ref{eq:V1}) (solid line), and (\ref{eq:fullV}) with $\kappa=1/2$ (dotted), $\kappa=\ln(1+\sqrt2)$ (dashed) and $\kappa=1$ (dot-dashed). The gray lines end of inflation ($\epsilon=1$).}
\end{figure}

In Fig.~\ref{fig:potentials} we show a comparison between the model described by (\ref{eq:fullV})  and the one introduced in (\ref{eq:V1}). It is important to note that for small $\kappa$, both potentials become similar. Indeed, up to $\mathcal O (\kappa^3)$ terms,
\begin{equation}
V(\phi)\approx\frac{V_0}{V_0^{(1)}}\left[\left(\frac{\mpl}{\phi_0}\right)^4-\frac56\frac{\kappa^2\mpl^2}{\phi_0^2}\frac{\phi^2}{\phi_0^2}\right]V_1(\phi).
\end{equation} The zeroth order difference may be absorbed in the normalization of the potentials, so the potentials can be made almost  identical\footnote{To the extent that their overall normalizations are irrelevant, as is the case for all quantities derived from the slow roll parameters or the number of inflation $e$-folds.} for
\begin{equation}
    \phi\alt\sqrt{\frac65}\frac\mpl\kappa.
\end{equation}
Then, only a relatively large $\kappa$ would produce substantial differences between both models if we want to avoid using highly trans-Planckian fields. For larger fields, the differences are more obvious, as $V_1$ grows indefinitely while \begin{equation}\lim_{\phi\to\infty}V(\phi)=\frac{V_0}{4}{\rm csch^4\,}\left(\frac{\kappa\phi_0}{2\mpl}\right).\end{equation}

The slow roll parameters can be now obtained from (\ref{eq:y-der}) and (\ref{eq:f2}), and are given by
\begin{subequations}
\begin{equation}
\epsilon=
\frac{2\kappa^2 y^2(1-y^2)}{(y-\beta)^2},\label{eq:epsilon-model}
\end{equation}
\begin{equation}
\eta=\frac{2 \kappa ^2 y \left(1-2 y^2\right)}{y-\beta}\label{eq:eta-model}
\end{equation}
and
\begin{equation}
\xi=\frac{4 \kappa ^4 y^2 \left(y^2-1\right) \left(\beta +2 y \left(6 y^2-3 \beta  y-2\right)\right)}{(y-\beta )^3};\label{eq:xi-model}
\end{equation}\label{eq:srpar-model}
\end{subequations}
these can be easily combined with (\ref{eqs:indices}) and (\ref{eq:r}) to  explore the parameter space in terms of $n_s$, $r$ and $N$. The first step in that direction requires to find out the condition(s) for slow roll to end. The potential under consideration allows for two types of slow roll inflation: \emph{(i)}~one in which $\phi$ rolls down the potential towards a minimum at larger values, and \emph{(ii)}~one in which a large field rolls down the potential towards smaller values. The condition $\epsilon=1$ may be rewritten as the quartic polynomial equation
\begin{equation}
y^4+\frac{1-2\kappa^2}{2\kappa^2}y^2-\frac{\beta}{\kappa^2}y+\frac{\beta^2}{2\kappa^2}=0,
\end{equation} which has, in principle, four complex roots, which may be obtained following Ferrari's method~\cite{Tignol:2001}. For the polynomial $y^4+q y^2+r y+s$, the roots are found to be
\begin{equation}
    y=\sigma_1\sqrt{\frac u2}+\sigma_2\sqrt{-\frac u2-\frac q2 -\sigma_1\frac{r}{2\sqrt{2u}}},
\end{equation}
where $\sigma_1$ and $\sigma_2$ are two independent signs that generate the four solutions, and $u$ is a solution to the cubic equation $u^3+q u^2+(q^2/4-s)u-r^2/8=0$. This can be reduced by a change of variables $u=v-q/3$ to a depressed cubic equation $v^3-(s+q^2/12)v-(  2q^3+27r^2-72q s)/216=0$. Such equation, generally $v^3+a v+b=0$, has a solution given by Cardano's formula $v=\sqrt[3]{\mathcal A_+}+\sqrt[3]{\mathcal A_-}$, with $\mathcal A_\pm=-b/2\pm\sqrt{\Delta}$ and $\Delta=(a/3)^3+(b/2)^2$ which, reverting the changes of variables, is
\begin{equation}
\Delta=-\frac{\beta^2}{2^83^3\kappa^8}\left[32\kappa^2\beta^4+(1-8\kappa^2(5+4\kappa^2)\beta^2-(1-2\kappa^2)^3\right].
\end{equation}
It is clear that the nature of the solutions depends on the sign of $\Delta$, which is unconstrained. The lines $\Delta=0$, which separate both regions, may be solved explicitly for $\beta$. Out of the four possible solutions, only
\begin{equation}\beta_0(\kappa)=\frac{1}{8} \sqrt{-\frac{\left(16 \kappa ^2+1\right)^{3/2}}{\kappa ^2}+32 \kappa ^2-\frac{1}{\kappa ^2}+40}\end{equation} is in the $\beta\in[0,1]$ and $\kappa>0$ region. The previous equation determines a limit in the $\kappa>1/\sqrt2$ region, value below which $\beta_0$ becomes complex. Moreover, $\kappa=\frac12\sqrt{11+5\sqrt5}$ marks the point at which $\beta=1$. For $\beta>\beta_0(\kappa)$, $\Delta>0$. Then $\kappa<1/\sqrt2\Rightarrow\Delta>0$ and $\kappa>\frac12\sqrt{11+5\sqrt5}\Rightarrow\Delta<0$. Conversely, $\Delta<0\Rightarrow \kappa>1/\sqrt2$ and $\Delta>0\Rightarrow\kappa<\frac12\sqrt{11+5\sqrt5}$. For $1/\sqrt2<\kappa<\frac12\sqrt{11+5\sqrt5}$ the curve $\beta_0(\kappa)$ separates both regions.

\paragraph{$\Delta>0$ ---}
In this case, $\mathcal A_\pm$ is real, and $u$ may be written directly as $u=\sqrt[3]{\mathcal A_+}+\sqrt[3]{\mathcal A_-}-q/3$. It it possible to see that $\mathcal A_\pm>0$ over the region where $\Delta>0$. Moreover, even if $q>0$ somewhere, $u>0$ everywhere. If $\sigma_1=+1$, $y$ is real and $\sigma_2$ generate both solutions. The case $\sigma_1=-1$ corresponds to complex solutions in all range where $\Delta>0$. Then, the two solutions of interest here are given by
\begin{subequations}
\begin{equation}
    y_\pm=\sqrt{\frac u2}\pm\sqrt{-\frac u2-\frac q2 -\frac{r}{2\sqrt{2u}}} \label{eq:ysol-1}
\end{equation}
and
\begin{equation}
u=
\left(-\frac b2+\sqrt{\left(\frac b2\right)^2+\left(\frac a3\right)^3}\right)^\frac13+
\left(-\frac b2-\sqrt{\left(\frac b2\right)^2+\left(\frac a3\right)^3}\right)^\frac13
-\frac q3.\end{equation}
    \end{subequations}

    \paragraph{$\Delta<0$ ---}
In this case, the solution to the cubic equation contains complex terms, and it becomes convenient to define $\mathcal A_\pm=\mathcal A\exp(\pm i\theta)$, where $\mathcal A\equiv|\mathcal A_\pm|=\sqrt{b^2/4-\Delta}=\sqrt{-(a/3)^3}$ and $\theta=2\arctan[\sqrt{-\Delta}/(\mathcal A-b/2)]$, which allows to write $v=2\mathcal A^{\frac13}\cos(\theta/3)$ and see that it is explicitly real. Moreover, as $\theta$ ranges in $[0,\pi]$, $\cos(\theta/3)$ is not negative. This, together with the fact that $q<0$ if $\kappa>1/\sqrt2$ makes $u$ positive as well. In this region, though, both values of $\sigma_1$ generate real solutions. It is clear that $\sigma_1=\sigma_2=-1$ would yield a negative solution, irrelevant in this case. Further investigation reveals that the other solution with $\sigma_1=-1$ yields negative solutions as well. The other solutions are always contained in $[0,1]$. Then, the solutions in this case are the same $y_\pm$ defined in (\ref{eq:ysol-1}) where now $u$ is better expressed involving real numbers only as
\begin{equation}
u=2\sqrt{-\frac a3}\cos\left[\frac23\arctan\left(\frac{\sqrt{-(b/2)^2-(a/3)^3}}{\sqrt{-(a/3)^3}-b/2}\right)\right]-\frac q 3.
\end{equation}

The parameters $a$, $b$ and $q$ used in all solutions above are
\begin{subequations}
\begin{equation}
    a=-\frac{\beta ^2}{2 \kappa ^2}-\frac{\left(1-2 \kappa ^2\right)^2}{48 \kappa ^4},
\end{equation}
\begin{equation}
    b=\frac{\beta ^2 \left(1-2 \kappa ^2\right)}{12 \kappa ^4}-\frac{\beta ^2}{8 \kappa ^4}-\frac{\left(1-2 \kappa ^2\right)^3}{864 \kappa ^6},
\end{equation}
and
\begin{equation}
   q= \frac{1-2 \kappa ^2}{2 \kappa ^2}.
\end{equation}
\end{subequations}

\begin{figure}
    \centering
    \subfigure[\,Solution $y_+$ from (\ref{eq:ysol-1}).]{\includegraphics[width=.49\linewidth]{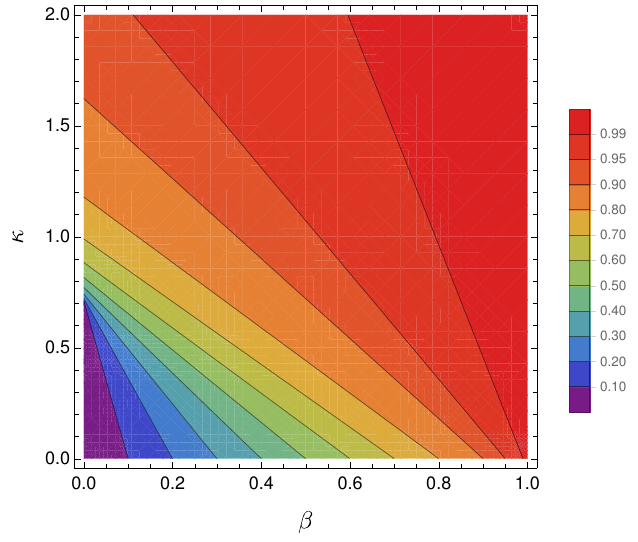}}
    \subfigure[\,Solution $y_-$ from (\ref{eq:ysol-1}).]{\includegraphics[width=.49\linewidth]{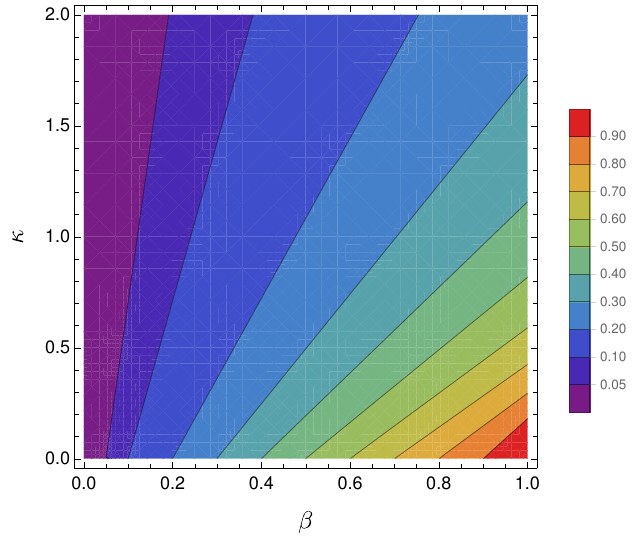}}
    \caption{Exploration of the solutions $y_\pm$ to the end of inflation condition $\epsilon=1$.}
    \label{fig:ysols}
\end{figure}

The solutions $y_\pm$ to the end of inflation equation $\epsilon=1$ are shown in Fig.~\ref{fig:ysols}. We recall here that $y_-$ corresponds to a solution for smaller $y$ (larger $\phi$), and $y_+$ to larger $y$ (smaller $\phi$). We define $y_\pm=\sech(\kappa\phi_\pm/\mpl)$.

We can now proceed with our analysis noting that, besides the values of $\beta$ and $\kappa$, which determine the end of inflation, there is still freedom in choosing the value of the field at the scale that corresponds to the experimental values. We call this $\phi_{*\pm}$ and define $\delta_\pm=\pm(\phi_\pm-\phi_{*\pm})/\mpl$. We recall that the distance Swampland conjecture demands that $\delta_\pm\alt\mathcal O(1)$. In terms of our model, when we are considering large fields (minus signs above), the value of $\phi_{*-}$ is unbounded and $\delta_-$ could take any positive value. Nevertheless, for small fields (plus signs above) $\delta_+$ is constrained so that $\phi_{*+}>0$, which means $\delta_+<\phi_+/\mpl$. All in all, the corresponding values for $y$ are
\begin{equation}
    y_{*\pm}=\sech\left(\vphantom{\sum}\asech\,y_\pm\mp\kappa\delta_\pm\right)\label{eq:ypmstar}
\end{equation}
where $\delta_+<\kappa^{-1}\asech\,y_+$.
A choice of values $(\beta,\kappa,\delta)$ and a branch (large or small $\phi)$ specifies the model completely. In specifically, the values for $y_\pm$ and $y_{*\pm}$ described above can be used to calculate the slow roll functions at $y_{*\pm}$ and, subsequently, the values of $n_s$ and $r$; and the number of $e$-folds between $y_{*\pm}$ and $y_\pm$.  Finally, we also study the number of inflation $e$-folds produced corresponding to the different parameters, which can be obtained from (\ref{eq:nefolds}) to be
\begin{equation}
N=\frac{
2 \beta  \left(\frac{1}{y_{*\pm}}-\frac{1}{y_{\pm}}\right)
+(1-\beta) \ln \left(\frac{1-y_{\pm}}{1-y_{*\pm}}\right)
+(1+\beta) \ln \left(\frac{1+y_{\pm}}{1+y_{*\pm}}\right)
-2 \ln \left(\frac{y_{\pm}}{y_{*\pm}}\right)}{4 \kappa ^2}.\label{eq:nefolds-model}
\end{equation}

\begin{figure}
    \centering
    \includegraphics[width=0.9\linewidth]{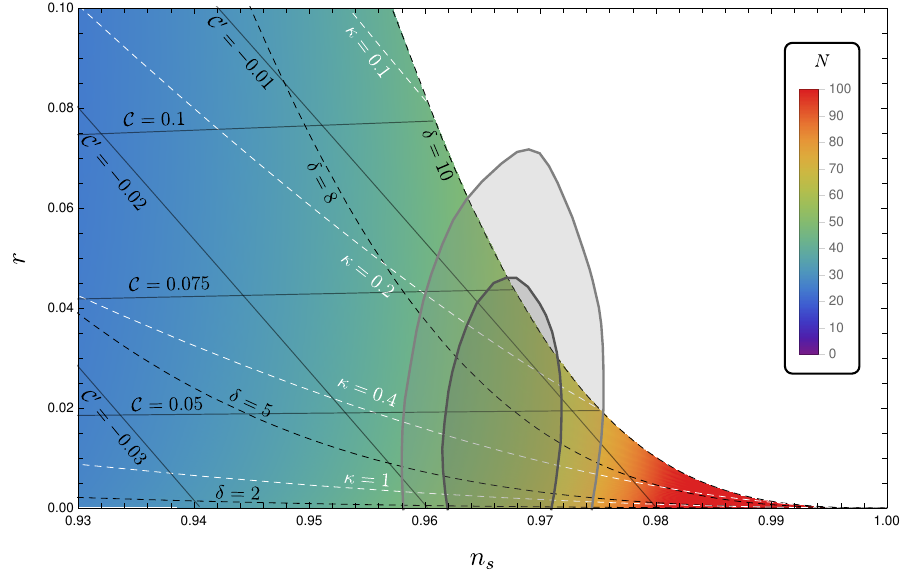}
    \caption{Exploration of the $(\kappa,\delta)$ parameter space for a fixed $\beta=0.5$. The dashed curves in black (white) are for constant $\delta$ ($\kappa$). The solid lines contain the relations between $(\mathcal C,\mathcal C')$ and $(r_s,r)$ for the model (i.e. for the actual $\xi$ obtained from (\ref{eq:xi-model})). The experimental bounds are as in Fig.~\ref{fig:nlo}.}
    \label{fig:final-large}
\end{figure}

Our results for the large field solution (corresponding to $y_-$) with $\beta=0.5$ are encapsulated in Fig.~\ref{fig:final-large}. The results show a mixed degree of compatibility between the Swampland conjectures and experimental data for the $S$-dual potential. The model itself can easily accommodate the experimental constrains for some region of the parameter space (namely $\kappa \agt 1$ and $\delta\alt1$), as $n_s$, $r$, and $50 \alt N \alt 60$ are all reproducible.  We can further study the compatibility between the de Sitter conjecture, the distance conjecture, and the experimental results. It is clearly visible how the bound on $\mathcal C$ from the de Sitter conjecture and the requirement on $\Delta \phi$ from the distance conjecture are in tension, as values of $\mathcal C\sim\mathcal O(1)$ even for the $95\%\,\mathrm{CL}$ lower limit on $n_s$ at $0.959$ require $\delta \gg 1$. On the other hand, the de Sitter bound on $\mathcal C'$ and the distance conjecture set bounds that get softer \emph{in the same direction} of decreasing $r$. In this case, $\mathcal C'$ is constrained by the data at the $95\%\,\mathrm{CL}$ lower limit on $n_s$ to $\mathcal C'\agt -0.02$. Despite this experimental constraint being much stronger than the $\mathcal C \alt 0.09$, there is no strong incompatibility with the distance conjecture.

\begin{figure}\centering\includegraphics[width=.55\linewidth]{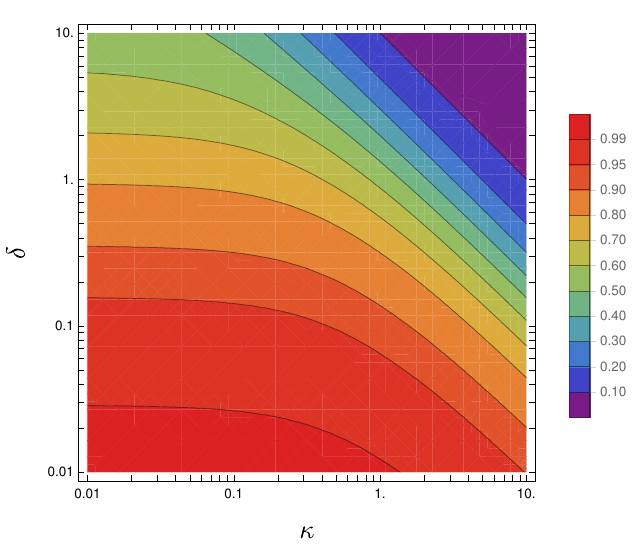}\caption{Value of $N\mathcal C/\delta$, bound to be lower than one by (\ref{eq:delta_bound}), as a function of $(\kappa,\delta)$ for $\beta=0.5$.}\label{fig:lyth}\end{figure}

As a final remark, we study the strength of Lyth's bound (\ref{eq:delta_bound}) on the current model, i.e. to which extent the model saturates such bound. In Fig.~\ref{fig:lyth} we show the value of $N\mathcal C/\delta$ as obtained from (\ref{eq:nefolds-model}) as a function of $\kappa$ and $\delta$, which is bound above by $1$ by means of Lyth's bound. It is visible that only for small values of $\delta$ Lyth's bound is saturated.

  \section{Ambiguity in slow-roll parameter definitions and impact on the Swampland conjectures}\label{sec:4-new}
  It is common in the literature to observe two different definitions of the slow-roll parameters, one defined in terms of the Hubble parameter $H$, and other in terms of the potential $V$; we have used the latter in previous sections. We have seen that the slow-roll parameters of single field inflation defined by $V$ are in tension with the Swampland conjectures. An interesting question we explore in this section is whether  these two choices of parameters differ in a significant way, so that the tension with the Swampland conjectures can be reduced.

  Accelerated expansion occurs as long as $\ddot a>0$, and, since
  \begin{equation}
    \frac{\ddot a}{a H^2}=1+\frac{\dot H}{H^2},
  \end{equation}
  that condition may be rewritten as $-\dot H/H^2<1$.
  The slow-roll limit means that $H$ is constant, as this is the only way to support exponential expansion with $a=\exp(Ht)$. The slow-roll regime may be considered as that in which $H$ changes slowly, which is what motivates the definition of the dimensionless slow roll parameter as
  \begin{equation}
    \epsilon_H=-\frac{\dot H}{H^2},
  \end{equation}
  for which $\epsilon_H<1$ means accelerated expansion, $\epsilon\ll1$ means slow-roll expansion, and $\epsilon_H=0$ means exponential expansion. Using the Friedmann equation \begin{equation}H^2=\frac{1}{3\mpl^2}\left(V+\frac12\dot\phi^2\right)\end{equation}
  and the equation of motion (\ref{eq:eom}), this may be rewritten as
  \begin{equation}
  \epsilon_H=3\frac{\dot\phi^2/2}{V+\dot\phi^2/2}=2\mpl^2\left(\frac{H_\phi}{H}\right)^2,\label{eq:app-eh}
  \end{equation}
  where we have made use of the relation $\dot\phi^2=-2\mpl^2 \dot H$. In order to connect this to the $V$-parameters, we write
  \begin{equation}
\frac{H_\phi}{H}=-\frac{3H\dot\phi}{6\mpl^2H^2}=\frac{V_\phi+\ddot\phi}{6\mpl^2H^2},
  \end{equation}
  and
  \begin{equation}
\epsilon_H=\frac{\mpl^2}{2}\left(\frac{V_\phi+\ddot\phi}{ V+\dot\phi^2/2}\right)^2.
  \end{equation}

  The slow-roll condition $\epsilon_H\ll1$ directly implies that $V\gg\dot\phi^2/2$, in which case
  \begin{equation}
\epsilon_H\approx\frac{\mpl^2}{2}\left(\frac{V_\phi+\ddot\phi}{V}\right)^2.
  \end{equation}
If one further imposes the condition that $|\ddot\phi|\ll|V_\phi|$, the approximation
\begin{equation}
  \epsilon_H\approx \frac{\mpl^2}{2}\left(\frac{V_\phi}{V}\right)^2 \label{eq:app-ehapprox}
\end{equation}
is valid. This motivates the definition of the $V$-parameter as
  \begin{equation}
  \epsilon_V\equiv\frac{\mpl^2}{2}\left(\frac{V_\phi}{V}\right)^2,
  \end{equation}
The question of whether $\epsilon_H$ and $\epsilon_V$ may be approximately equal depends on whether the two approximations used to derive (\ref{eq:app-ehapprox}) are simultaneously satisfied. A glance at (\ref{eq:app-eh}) suggests that they may not always be, as in the limit $\dot\phi\to0$, $\epsilon_H\to0$ while $\epsilon_V$ may take any finite value. Moreover, one can rewrite the equation of motion as
\begin{equation}
(V_\phi+\ddot\phi)^2=\frac{3}{\mpl^2}\dot\phi^2(V+\dot\phi^2/2)
\end{equation}
to see that a condition on the smallness of $\dot\phi^2$ doesn't guarantee the smallness of $\ddot\phi$ with respect to $V_\phi$ unless $V_\phi$ is itself small. We conclude that, in general, both conditions must be separately satisfied to guarantee the similarity between $\epsilon_H$ and $\epsilon_V$. A more comprehensive study of the differences between both parameters (as well as the second order ones, $\eta_H$ and $\eta_V$), can be found in \cite{Liddle:1994}. Here we highlight only the aspects relevant for our discussion.

  Given a specific potential $V(\phi)$, one can obtain a solution $\phi(t)$ to the equation of motion, subject to the initial conditions $\phi(t_0)=\phi_0$ and $\dot\phi(t_0)=\dot\phi_0$. This makes the difference between $\epsilon_V$ and $\epsilon_H$  explicit, since at $t=t_0$ the former depends only on $\phi_0$ while the latter depends on both $\phi_0$ and $\dot\phi_0$. Therefore, the equality or similarity between $\epsilon_V$ and $\epsilon_H$ is a matter of a handpicked pair $(\phi_0,\dot\phi_0)$ that would guarantee both $\dot\phi^2/2\ll V $ and $|\ddot\phi|\ll V_\phi$.

  This makes clear that the end of inflation condition $\epsilon_H=1$ would yield different results than $\epsilon_V=1$. While the latter condition is the most commonly used, and is simpler to evaluate due to its sole dependence on the shape of the potential, it is the former condition that must be satisfied exactly, since it depends on the full solution of the scalar field equation of motion.

A judicious choice of initial conditions on the field and its derivative at the time at which the scale $k_*$ crosses the horizon should be able to accommodate multiple values of $\mathcal C$ or $\mathcal C'$, potentially reducing the tensions with the Swampland conjectures, while remaining in the $H$-dictated slow-roll regime. Nevertheless, keeping the de Sitter conjecture and the observed number of inflation $e$-folds under control is not guaranteed in this situation. A full study like the one presented in Sec.~\ref{sec:3}, adding these initial conditions, should be considered if one aims to characterize the complete parameter space. Nevertheless, we leave that for future work, as the increase in computational complexity escapes the aim of this paper. Here, instead, we choose initial conditions that optimize the comparison between the $H$-parameters and the $V$-parameters, rather than the generality of the study.

%Below, we proceed to study to which extent the full account given by the condition $\epsilon_H=1$ yields significantly different results from the condition $\epsilon_V=1$ used in previous sections.

  To remove part of the ambiguity caused by the freedom of choice in the initial conditions, we consider $t_*$ (the time at which the scale $k_*$ crosses the horizon) as the starting point for the solution to the equation of motion. In order to reduce the number of quantities affected by the choice of parameters, we choose to leave the observable values of $n_s$ and $r$ unaffected. Since these values depend on $\epsilon$ and $\eta$ at $t_*$, fixing the initial conditions on $\phi$ such that $\epsilon_H(t_*)=\epsilon_V(t_*)$ allows to remove any effect of this choice on them. This is just an operational perspective that should allow us to compare the differences that $\epsilon_H$ and $\epsilon_V$ have only in regard to the other observable, $N$. This condition amounts to
  \begin{equation}
  \dot\phi^2(t_*)=\frac{2V(\phi_*)}{3/\epsilon_V(\phi_*)-1},
  \end{equation}
  which may be rewritten in terms of $y$ as\footnote{Only the large $\phi$ (small $y$) solution (the negative sign in the $y_\pm$ notation) is considered here, as we deemed it to be the interesting case. Otherwise, the initial derivative should be negative.}
  \begin{equation}
    \dot y(t_*)=\frac{\sqrt{{V_0}}}{\mpl}\frac{ 2 \kappa ^2 {y_0}^2(\beta-y_0)\left(1-{y_0}^2\right)}{(1-\beta ) \sqrt{3 ({y_0}-\beta )^2-2 \kappa ^2 {y_0}^2 \left(1-{y_0}^2\right)}}.
  \end{equation}

  In a similar manner as we proceeded before, we start by fixing the model parameters $(\beta,\kappa)$, and finding the end of inflation using the $\epsilon_V(y_{\mathrm{e},V})=1$ condition, and the value of $y_*$ via (\ref{eq:ypmstar}), using a given value of $\delta$, named here $\delta_V$. With both $y_*$ and $y_{\mathrm{e},V}$ we can find the value of $N_V$ using (\ref{eq:nefolds-model}). To quantify the difference between the choice of $\epsilon_V$ and of $\epsilon_H$, we calculate the true end of inflation through the $\epsilon_H(y_{\mathrm{e},H})=1$ condition, which provides a true value for $\delta$, as $\delta_H=\kappa^{-1}(\asech\,y_*-\asech\,y_{\mathrm e,H})$, and a true number of $e$-folds $N_H$ as
  \begin{equation}
  N_H=-\frac1\mpl\int_{\phi_*}^{\phi_e}\frac{d\phi}{\sqrt{2\epsilon_H}}=\frac1\kappa\int_{t_*}^{t_{\mathrm e,H}}\frac{\dot y}{y\sqrt{1-y^2}\sqrt{2\epsilon_H}}dt,
  \end{equation}
  where $t_{\mathrm e,H}$ is the true time at which inflation ends.

\begin{figure}[h!]
      \centering
      \subfigure[\label{fig:app-fig-a}\,$N$ as a function of $\delta$ for the $\epsilon_H=1$ (solid) and $\epsilon_V=1$ (dashed) conditions.]{\includegraphics[height=6.5cm]{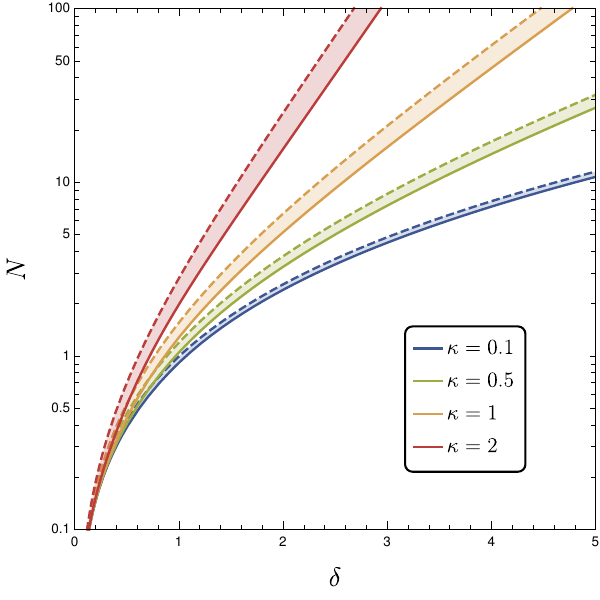}}
  \quad    \subfigure[\label{fig:app-fig-b}\,Underestimation of $N$ with $\epsilon_V=1$ relative to $\epsilon_H=1$.]{\includegraphics[height=6.6cm]{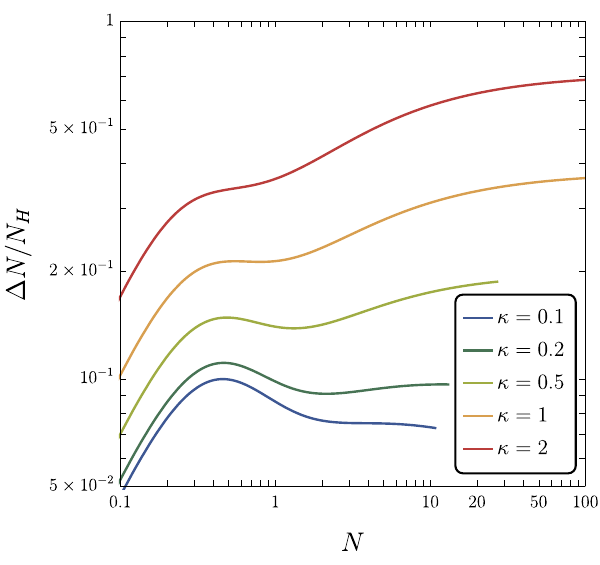}}
      \caption{Comparison of the two end of inflation conditions, $\epsilon_H=1$ and $\epsilon_V=1$, regarding their effect on the number of inflation $e$-folds and the parameter $\delta$.}
      \label{fig:app-fig}
  \end{figure}

  In Fig.~\ref{fig:app-fig-a} we show the relation between $N_V$ and $\delta_V$, and $N_H$ and $\delta_H$. While a full analysis similar to that presented in Fig.~\ref{fig:final-large} might be the only way to fully understand the relevance of the parameter set choice, here it is visible how the difference $\Delta N\equiv N_H-N_V$ grows both with $\kappa$ and $\delta$. We can see in Fig.~\ref{fig:app-fig-b} that for small values of $\kappa$, which are of more interest in the study of the de Sitter conjecture on $\mathcal C$, $\Delta N$ is small enough to make it irrelevant, and no significant difference would be expected in that front. Regarding $\mathcal C'$, the $\mathcal C'\agt-0.02$ experimental bound may be accommodated a bit easier regarding the number of $e$-folds, as in that region the constraint $\epsilon_V=1$ is underestimating the number of $e$-folds by a few percent points of its true value.

  Nevertheless, despite the minor changes introduced in relation to the Swampland conjectures, Fig.~\ref{fig:app-fig-b} makes clear that the $\epsilon_V=1$ condition might end inflation too early, producing considerable underestimations of the actual number of $e$-folds.

  We want to remind the reader that these results are obtained for $\beta=0.5$, as was the case with the results presented in and after Fig.~\ref{fig:final-large}. It must also be clarified that the information presented in Fig.~\ref{fig:app-fig-a} and Fig.~\ref{fig:app-fig-b} is not in contradiction, unlike it may seem. Fig.~\ref{fig:app-fig-a} presents the curves $N_H(\delta_H)$ and $N_V(\delta_V)$, so the horizontal axis is not the same variable, and makes it seem that for a single $\delta$, the $V$-based condition overestimates the number of $e$-folds. Nevertheless, in Fig.~\ref{fig:app-fig-b} we show the curves evaluated at the same value of $\delta$. This is therefore comparing the two parameter choices for a fixed value of the field excursion $\Delta\phi$. Under this circumstance, it is clearly seen that the $V$-based conditions produce an underestimation of the number of inflation $e$-folds with respect to the $H$-based ones.

\section{Conclusions}
\label{sec:4}

We have analyzed the most general form of single-field $S$-dual inflationary
potentials at NLO in slow-roll parameters within the context of the
Swampland program and confronted model predictions with experiment. We
have found that to accommodate the 95\% CL limit on $r <
0.068$ form  BICEP2/Keck Array + Planck  + BAO
data~\cite{Ade:2018gkx,Akrami:2018odb} we require $c \sim {\cal
  O} (10^{-1})$. This requirement is in tension with the de Sitter
conjecture. However, in the spirit of~\cite{Dias:2018ngv}, we can
adopt a conservative approach and regard the de Sitter conjecture as a
parametric constraint where the inequality (\ref{dS1}) holds, but the
number $c$ may not be strictly ${\cal O}(1)$. Indeed, it is easy to
established a mass hierarchy between the lightest moduli field and and
inflaton to accommodate $c \sim {\cal
  O}(10^{-1})$~\cite{Dias:2018ngv}. From this viewpoint,
constraints on inflation can then be used to constrain $c$. Still, as
we have shown in Fig.~\ref{fig:final-large}  to accommodate $c \sim {\cal
  O}(10^{-1})$ a $\delta \sim {\cal O}(10)$ would be required. To be
able to match such a large  value of $\delta$ we must explore the
subtleties of  the distance conjecture, which asserts that for any
infinite field distance limit, an infinite tower of states becomes
exponentially light, and therefore EFTs are only valid for finite scalar field variations~\cite{Ooguri:2006in,Klaewer:2016kiy, Ooguri:2018wrx,Grimm:2018ohb,Heidenreich:2018kpg}.  This in turn implies a quantum gravity cutoff associated to the infinite
tower of states, decreasing exponentially in terms of the proper field
distance, $\Lambda_{\rm QG} = \Lambda_{\rm self} \ e^{-\lambda \, \Delta \phi}$,
  where $\Lambda_{\rm QG}$ is the quantum gravity cutoff,
  $\Lambda_{\rm self}$ is the
cutoff of the EFT, and $\lambda$ is argued to be of order unity in Planck
units (see, however,~\cite{Andriot:2020lea,Gendler:2020dfp}). Now, since $\Lambda_{\rm self} \leq M_{\rm Pl}$ we have $\Delta \phi \leq
\lambda^{-1}
    \ln (M_{\rm Pl}/\Lambda_{\rm self})$, which indicates
that the maximum field variation actually depends on the cutoff of
the EFT~\cite{vanBeest:2021lhn}. We know that
for EFT to
describe inflation, its cutoff must be above the Hubble scale, i.e. $\Lambda_{\rm self} > H$. If we adopt the conservative bound $\Lambda_{\rm self} \sim H$,
then $\Delta \phi \alt 10~M_{\rm Pl}$~\cite{Scalisi:2018eaz}. Needless
to say, it should  be stressed that
the EFT will likely break down (or at least get sensitive to the
infinite tower) before the mass of the first state becomes of order
Hubble, so the constraints might be stronger than those derived from
the assumption $\Lambda_{\rm self} \sim H$. Next-generation CMB
satellites searching for primordial B-modes (e.g.
PIXIE~\cite{Khatri:2013dha}, CORE~\cite{Finelli:2016cyd}, and  LiteBIRD~\cite{Hazumi:2019lys})  will
reach a 95\%CL sensitivity of $r < 0.002$. This will allow discrimination
between small-field $\Delta \phi< M_{\rm Pl}$ and large-field $\Delta
\phi > M_{\rm Pl}$ inflationary models, and will provide a final verdict for the ideas presented and discussed in this paper.

As a final remark, it would be interesting to study the full parameter space using the $H$-parameters introduced in Sec.~\ref{sec:4-new} rather than the $V$-parameters. As stated there, the increase in the number free parameters would make it more feasible to reduce the tension with the Swampland conjectures. An analysis like the one presented here in which the $H$-parameters are used in full is left for future work.

\acknowledgments{The work of L.A.A. and J.F.S. is supported by the
  by the U.S. National Science Foundation (NSF Grant PHY-1620661) and
  the National Aeronautics and Space Administration (NASA Grant
  80NSSC18K0464). The research of I.A. was partially performed as International professor of the Francqui Foundation, Belgium. The work of D.L. is
  supported by the Origins Excellence Cluster.  Any opinions, findings, and
  conclusions or recommendations expressed in this material are those
  of the authors and do not necessarily reflect the views of the NSF
  or NASA.}

\end{document}